\documentclass[11pt, a4paper, final, oneside, onecolumn]{article}
\pdfoutput=1

\usepackage[a4paper, margin=1in]{geometry}
\usepackage{ragged2e}
\usepackage{graphicx}
\graphicspath{ {/Users/apple/Documents/paper} }
\usepackage{algorithm}
\usepackage{algpseudocode}
\usepackage{pgfplots}

\begin{document}

	\title{A Constant Optimization of the Binary Indexed Tree Query Operation}
	\author{Anubhav Baweja}
	
	\maketitle

	\begin{abstract}
		There are several data structures which can calculate the prefix sums of an array efficiently, while handling point updates on the array, such as Segment Trees and Binary Indexed Trees (BIT). Both these data structures can handle the these two operations (query and update) in $O(\log{n})$ time. In this paper, we present a data structure similar to the BIT, but with an even smaller constant. To do this, we use Zeckendorf's Theorem, a property of the Fibonacci sequence of numbers. The new data structure achieves the same complexity of $O(\log{n})$, but requires about $\log_{\phi^{2}} n$ computations for the Query Operation as opposed to the $\log_{2} n$ computations required for a BIT Query Operation in the worst case.
	\end{abstract}
	 
	\section{Problem Motivation}
	
	A Prefix Sum is defined as the sum of the first $n$ elements of an array, where $ 1 \leq n \leq size(array) $. The problem can be traditionally solved on an array $arr$ by creating a prefix sum array $pre$ such that
	\begin{equation}
		pre[1] = arr[1]
	\end{equation}
	\begin{equation}
		pre[i] = pre[i-1] + arr[i] 
	\end{equation}
	(The above equations follow 1-based indexing)
	\\\\
	The Query Operation is defined as calculating the Prefix Sum of any index. The Update Operation is defined as assigning a new value to any index in the $arr$ array. 
	\\\\
	It is easy to see that the above $pre$ array can handle the Query Operation in $O(1)$ time, since it only requires a single memory call. However, the Update Operation is highly inefficient without the use of any data structure, taking $O(n)$ time in the worst case (when updating the last element in the array). 
	\\\\
	The Segment Tree and the Binary Indexed Tree (BIT) \cite{fenwick1994new} are two structures which can handle both the Query and the Update Operation in $O(\log{n})$ time. However, a BIT is much more efficient than a Segment Tree due to a smaller constant. In this paper, we present an alternative to the BIT, which we will call the Fibonacci Indexed Tree (FIT), for the sake of convenience. It can be shown that in the worst case, FIT takes about $\log_{\phi^2} n$ computations for the Query Operation and about $\log_{\phi} n)$ computations for the Update Operation. 
	\\\\
	The problem of handling the above two operations simultaneously on a collection of data is an important one. It is used to solve several problems such as the Line-of-Sight Problem and is used in the implementation of multiple algorithms such as Radix Sort, lexical comparison of strings and Arithmetic Coding for data compression \cite{blelloch1990prefix} \cite{witten1987arithmetic}. When extended to two dimensions, Prefix Sums (the sum of all elements in a prefix rectangle) can be used in image processing and geographical information systems \cite{samet1984geographic}.

	\section{Preliminaries}
	
	Before introducing the data structure, we will first tackle the few definitions and concepts which are required to understand its mechanism. We will be discussing the following topics:
	\begin{itemize}
		\item Zeckendorf's Theorem \cite{zeckendorf1972representation}
		\item Fibonacci Coding
		\item Least Significant Used Fibonacci
		\item Mechanism of the Binary Indexed Tree
	\end{itemize}
	
	\subsection{Zeckendorf's Theorem}
	
	This theorem states that every positive integer can be expressed as the sum of distinct non-consecutive terms of the Fibonacci Sequence of numbers.
	Example: $46$ can expressed as $46 = 34 + 8 + 3 + 1$. This is called the Zeckendorf Representation of the number. The theorem also states that there exists only one Zeckendorf Representation of every positive integer.
	\\\\
	The Zeckendorf Representation of a number can be obtained by using a Greedy Algorithm. At every step, simply take the largest Fibonacci number smaller than the required number and subtract it. Now, repeat the step until a Fibonacci number is obtained. All the Fibonacci Numbers that were subtracted and the number finally obtained together make up the Zeckendorf Representation.

	\subsection{Fibonacci Coding}
	
	An alternate way to write the Zeckendorf Representation of a number is 'Fibonacci Coding'. For a number $N$, we define the Fibonacci Coding of the number as follows:
	
	\begin{equation}
		N = \sum_{i=0}^{k-1} d(i)F(i+2) 
	\end{equation}
	
	where $F(i)$ is the $i$th Fibonacci number starting from 1, $d(i)$ is the $i$th digit in the Fibonacci Coding which is either one or zero, and $k$ is the number of digits in the Fibonacci Coding. This means that the $i$th digit of the Fibonacci Coding is $1$ if the $(i+2)$th Fibonacci number is part of its Zeckendorf Representation, and $0$ otherwise. 
	\\\\
	Example, $46$ is written as $46 = 34 + 8 + 3 + 1$. The following are the values for $F(i)$ and $d(i)$:
	
	\begin{table}[ht]
		\centering
		\begin{tabular}{|l|l|l|l|l|l|l|l|l|}
			\hline
			i    & 0 & 1 & 2 & 3 & 4 & 5  & 6  & 7  \\ \hline
			F(i) & 1 & 2 & 3 & 5 & 8 & 13 & 21 & 34 \\ \hline
			d(i) & 1 & 0 & 1 & 0 & 1 & 0  & 0  & 1  \\ \hline
		\end{tabular}
	\end{table}
	
	Thus, the Fibonacci Coding of $46$ is given as $10101001$.
	\\\\
	However, note that the most significant Fibonacci number is positioned to the right (the place value of digits increases from left to right). This is opposite to what we observe in the decimal or binary representation of numbers, where the place value of the digit increases from right to left. Thus, for the purposes of this paper, we will use the reverse Fibonacci Coding. For the sake of convenience, we will call it the Fibonacci Coding of the number. Therefore, the Fibonacci Coding of $46$ is given as $10010101$.
	
	\subsection{Least Significant Used Fibonacci}
	Formally, the Least Significant Bit of a positive integer is defined as the bit position in the binary representation of the number, which gives its units value, that is, determines whether the number is odd or even. However, for the purposes of this paper, we will define the "Least Significant Used Bit" (LSUB): the smallest power of two which is part of the Binary Representation. For instance, the Binary Representation of $40$ is $101000$. Therefore, the LSUB of $40$ is $8$. 
	\\\\
	Similarly, we can also define the "Least Significant Used Fibonacci" (LSUF) of a number as the smallest Fibonacci Number which occurs in the Fibonacci Coding of a number. For instance, the Fibonacci Coding of $45$ is $10010100$. Thus, the LSUF of $45$ is $3$.
	\\\\
	In the following sections, we will use $lsub(x)$ and $lsuf(x)$ to denote the the least significant used bit and fibonacci of $x$ respectively.
	
	\subsection{Mechanism of the Binary Indexed Tree}
	Before moving on to the working of the Fibonacci Indexed Tree, we must be familiar with how a Binary Indexed Tree calculates prefix sums. Let's define an array $bit$, which stores the Binary Indexed Tree. Each element of the $bit$ array stores the sum of a certain range in the $arr$ array (the original array which contains our data). 
	
	\begin{figure}[ht]
		\includegraphics[width = \textwidth]{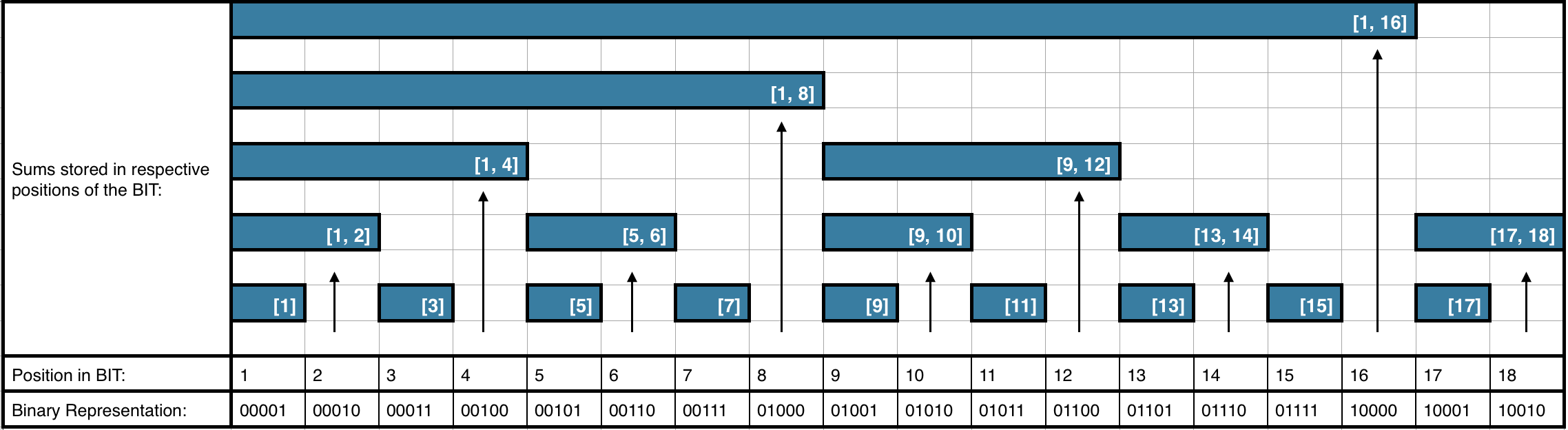}
		\caption{How sums are stored in the $bit$ array}
		\label{fig:fig1}
	\end{figure}
	
	Figure \ref{fig:fig1} shows hows sums are stored in the $bit$ array. For instance, $bit[12]$ stores the sum of elements from $arr[9]$ to $arr[12]$. The range of values that $bit[x]$ stores is given as follows:
	\begin{equation}
		bit[x] = \sum_{i = y}^{x} arr[i] \quad ; \quad y = x-lsub(x)+1
	\end{equation}
	
	\subsubsection{Query Operation}
	In order to calculate the prefix sum ending at $n$, we follow the below algorithm:
	\\
	\begin{algorithm}[H]
   		\caption{BIT Query Operation}
		\begin{algorithmic}[1]
  			\Function{queryb}{$x$}
    			\State $sum := 0$
    			\While{$x > 0$}
        			\State $sum := sum + bit[x]$
        			\State $x := x - lsub(x)$
    			\EndWhile
    			\State \Return $sum$
    		\EndFunction
		\end{algorithmic}
	\end{algorithm}
	
	It can be seen that the loop in the above code runs the same number of times as the number of ones in the Binary Representation of $n$, since each iteration turns the LSUB into a zero. Now, the Binary Representation of $n$ contains at most $\log_{2} n$ digits. Therefore, in the worst case, the loop runs $\log_{2} n$ times (when $n = 2^k - 1$ for some positive integer $k$).
	
	\subsubsection{Update Operation}
	As stated earlier, the advantage of using a Binary Indexed Tree (BIT) over the $pre$ array is that the BIT can support the Update Operation, that is, we can change the value of an element in the $arr$ array in $O(\log{n})$ time, rather than $O(n)$ time. To update an element and set its value to $c$, the following algorithm must be used:
	\\
	\begin{algorithm}[H]
   		\caption{BIT Update Operation}
		\begin{algorithmic}[1]
  			\Function{updateb}{$x, c$}
    			\State $temp := c - arr[x]$
    			\State $n := $ size of $arr$
    			\While{$x \leq n$}
        			\State $bit[x] := bit[x] + temp$
        			\State $x := x + lsub(x)$
    			\EndWhile
    		\EndFunction
		\end{algorithmic}
	\end{algorithm}
	
	\section{The Fibonacci Indexed Tree}
	
	The essential mechanism of the Fibonacci Indexed Tree (FIT) is the same as that of the BIT: we declare an array $fit$, which contains the data structure. As before, $fit[i]$ stores the sum of a certain range of values in the original $arr$ array, where our data is stored.
	
	\begin{figure}[H]
		\includegraphics[width = \textwidth]{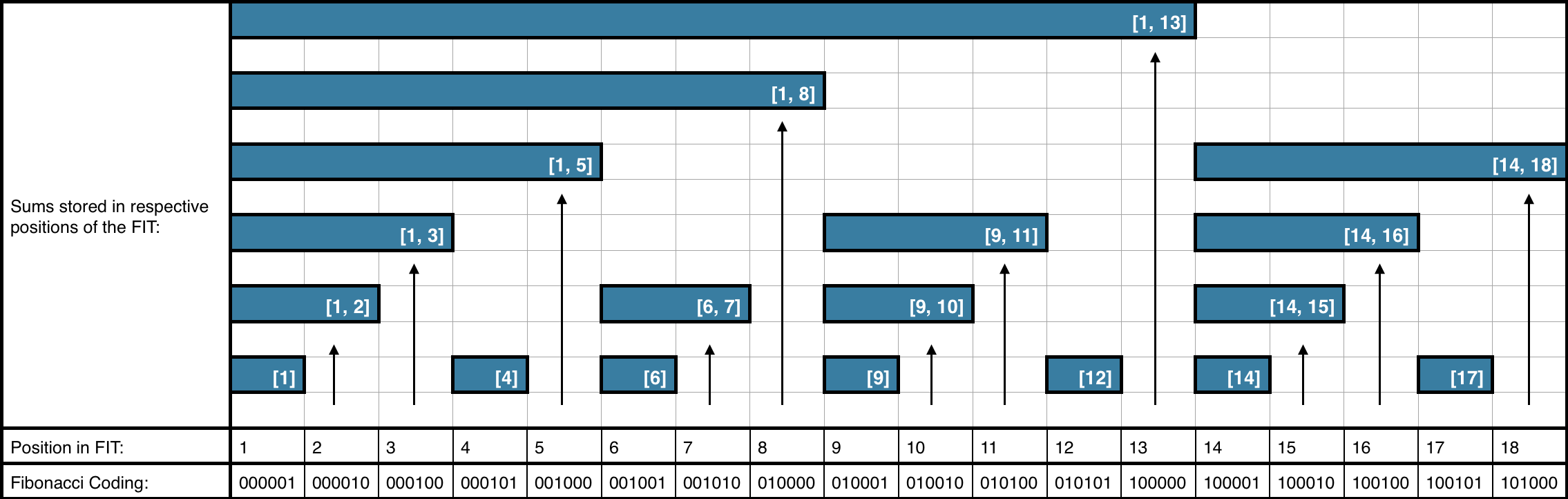}
		\caption{How sums are stored in the $bit$ array}
		\label{fig:fig2}
	\end{figure}
	
	Figure \ref{fig:fig2} shows hows sums are stored in the $fit$ array. For instance, $fit[11]$ stores the sum of elements from $arr[9]$ to $arr[11]$. The range of values that $fit[x]$ stores is given as follows:
	\begin{equation}
		fit[x] = \sum_{i = y}^{x} arr[i] \quad ; \quad y = x-lsuf(x)+1
	\end{equation}
	
	Notice that the only thing different in Equation (5) from Equation (4) is the use of the $lsuf$ function instead of the $lsub$ function. We will also use the following equation to prove the time complexities of the Query and the Update Operation:
	\begin{equation}
		\lim_{n \to \infty} \frac{F_{n+1}}{F_{n}} = \phi
	\end{equation}
	
	\subsection{Query Operation}
	
	\begin{algorithm}[H]
   		\caption{FIT Query Operation}
		\begin{algorithmic}[1]
  			\Function{queryf}{$x$}
    			\State $sum := 0$
    			\While{$x > 0$}
        			\State $sum := sum + fit[x]$
        			\State $x := x - lsuf(x)$
    			\EndWhile
    			\State \Return $sum$
    		\EndFunction
		\end{algorithmic}
	\end{algorithm}
	
	The loop in the above code runs the same number of times as the number of ones in the Fibonacci Coding of $x$. There are approximately $\log_{\phi} x$ Fibonacci numbers less than $x$. This follows from Equation (6). Thus, there are $\log_{\phi} x$ digits in the Fibonacci Coding of $x$. This would imply that in the worst case, approximately $log_{\phi} x$ computations would be required (when $x = F(k) - 1$, where $F(k)$ is the $k$th Fibonacci Number). However, that is not the case. 
	\\\\
	No two consecutive positions in the Fibonacci Coding can be $1$, otherwise the two consecutive Fibonacci Numbers will add up to form the next Fibonacci Number, thus eliminating both of the original numbers. Thus, in the worst case, only half of the digits in the Fibonacci Coding are ones. Therefore, the number of computations required becomes: $0.5 * \log_{\phi} n = \log_{\phi^2} n$.
	\\
	
	\begin{center}
		\begin{tikzpicture}
			\begin{semilogxaxis}[
			    axis lines = left,
			    xlabel = $n$,
			    ylabel = {$f(n)$},
			    legend pos = north west,
			    xmax = 1e6,
			    ymax = 20
			]
			
			\addplot [
			    domain=1:3e5, 
			    samples=20, 
			    color=red,
			]
			{ln(x)/ln((3 + sqrt(5))/2)};
			\addlegendentry{$\log_{\phi^2}{n}$}
			
			\addplot [
			    domain=1:3e5, 
			    samples=20, 
			    color=blue,
			]
			{ln(x)/ln(2)};
			\addlegendentry{$\log_{2}{n}$}
			
			\end{semilogxaxis}
		\end{tikzpicture}
	\end{center}
	
	\subsection{Update Operation} 
	
	\begin{algorithm}[H]
   		\caption{FIT Update Operation}
		\begin{algorithmic}[1]
  			\Function{updatef}{$x, c$}
    			\State $temp := c - arr[x]$
    			\State $n := $ size of $arr$
    			\While{$x \leq n$}
        			\State $fit[x] := fit[x] + temp$
        			\State $x := x + prefib(lsuf(x))$
    			\EndWhile
    		\EndFunction
		\end{algorithmic}
	\end{algorithm}

	In the above code, the $prefib(x)$ function returns the previous Fibonacci Number of $x$, where $x$ itself should be a Fibonacci Number. The motive of the above algorithm and the BIT Query Operation is the same: moving the Least Significant Used Bit/Fibonacci by (at least) one place to the left. By adding $prefib(lsuf(x))$, we will get a one in two consecutive positions of the Fibonacci Coding, which will add up to form the next Fibonacci number. This is slightly different from Algorithm 2: The BIT Update Operation, where we simply add the $lsub(x)$ to $x$. 
	\\\\
	The worst case occurs when $x = F(k) + 1$, for some positive integer $k$. This is because in each iteration of the loop, the position of the Least Significant Used Fibonacci shifts by a single place. Thus, approximately $\log_{\phi} x$ computations are required to perform the entire update operation, which is more than what we have in the BIT Update Operation.
	\\
	\begin{center}
		\begin{tikzpicture}
			\begin{semilogxaxis}[
			    axis lines = left,
			    xlabel = $n$,
			    ylabel = {$f(n)$},
			    legend pos = north west,
			    xmax = 1e6,
			    ymax = 30
			]
			
			\addplot [
			    domain=1:3e5, 
			    samples=20, 
			    color=red,
			]
			{ln(x)/ln((1 + sqrt(5))/2)};
			\addlegendentry{$\log_{\phi}{n}$}
			
			\addplot [
			    domain=1:3e5, 
			    samples=20, 
			    color=blue,
			]
			{ln(x)/ln(2)};
			\addlegendentry{$\log_{2}{n}$}
			 
			\end{semilogxaxis}
		\end{tikzpicture}
	\end{center}
	
	\section{Conclusion and Scope for Further Study}
	The proposed data structure is essentially a modification of the Binary Indexed Tree with a certain trade-off: faster queries for slower updates. For most practical purposes however, this trade-off is a favourable one. The most common use of Prefix sums is calculating "range sums", that is, for given positions $l$ and $r$, calculating the following:
	\begin{equation}
		\sum_{i=l}^{r} arr[i] = query(r) - query(l-1)
	\end{equation}
	
	As we can see, two $query$ operations are needed to obtain a single "range sum". On the other hand, updating a single element of the array only requires only one call to the $update$ function. Thus, having a faster $query$ function is preferable for most practical applications such as the Arithmetic Coding Algorithm for data compression \cite{witten1987arithmetic}.
	\\\\
	Like the Binary Indexed Tree, the Fibonacci Indexed Tree can also be extended to multiple dimensions \cite{mishra2013new} . In the 2D variant of the BIT, which is used in image processing and geographical information systems \cite{samet1984geographic}, $4$ $query$ operations are required to calculate the "2D range sum". Moreover the complexity for one $query$ operation is $O((\log{n})^2)$. Thus, switching to a Fibonacci Indexed Tree has true merit in this case. 
	\\\\
	
	\bibliography{ref.bib}
	\bibliographystyle{ieeetr}
	
\end{document}